\documentclass[preprint,preprintnumbers,showpacs,aps,amssymb]{revtex4}

\usepackage{bm}
\usepackage{epsf,epsfig,subfigure,axodraw,graphicx,amsmath,amssymb}

\def\lsim{\lower.7ex\hbox{$\;\stackrel{\textstyle<}{\sim}\;$}}

\def\Qem{{$Q_{\rm em}$}}

 \def\Z{{\bf Z}}

 \def\Z{{\bf Z}}

\begin{document}
\preprint{SNUTP 07-005}
\title{PVLAS experiment, star cooling and BBN constraints:
Possible interpretation with temperature dependent gauge symmetry
breaking
 }
\author{Jihn E.  Kim 
}
\address{Department of Physics and Astronomy and Center for Theoretical
 Physics, Seoul National University, Seoul 151-747, Korea }

\begin{abstract}
{It is known that the kinetic mixing of photon and another
U(1)$_{\rm ex}$ gauge boson can introduce millicharged particles.
Millicharged particles $f$ of mass 0.1 eV can explain the PVLAS
experiment. We suggest a temperature dependent gauge symmetry
breaking of U(1)$_{\rm ex}$ for this idea to be consistent with
astrophysical and cosmological constraints. }
 \keywords{PVLAS experiment, Millicharged particles,
 Temperature dependent potential}
\end{abstract}

 \pacs{14.80.-j, 95.30.Cq,  11.10.Wx}

 \maketitle

\begin{figure}[b]
\begin{center}
\begin{picture}(400,110)(0,0)
 {
 \Photon(40,90)(100,90){3}{9.5} \DashLine(100,90)(160,90){3}
 \Photon(100,90)(100,30){3}{9} \Text(100,30)[c]{$\bf \times$}
 \Text(37,90)[r]{$\gamma$} \Text(100, 20)[c]{$\bf B$}
 \Text(163,90)[l]{$a$}
 \Text(100,0)[c]{(a)}
 }
 {
 \Photon(240,90)(300,90){3}{10} \Line(300,90)(360,90)
 \Photon(240,30)(300,30){3}{10} \Line(300,90)(300,30)
 \Line(300,30)(360,30)
 \Text(240,30)[c]{$\bf \times$}
 \Text(237,90)[r]{$\gamma$} \Text(235,30)[r]{$\bf B$}
 \Text(363,90)[l]{$f$}
  \Text(363,30)[l]{$\overline{f}$}
 \Text(300,0)[c]{(b)}
 }
\end{picture}
\end{center}
\caption{Feynman diagrams leading to photon conversions in a
magnetic field.}\label{fig:FeynPVLAS}
\end{figure}
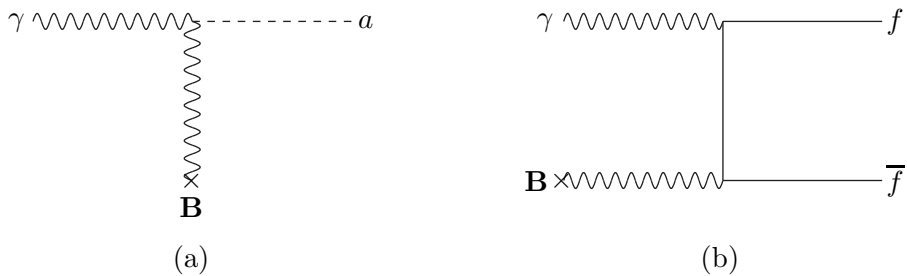

 The possibility of dichroism of vacuum in a magnetic field has been
 announced  by the PVLAS collaboration \cite{Zavattini}.
The vacuum magnetic dichroism arises from the absorption rate
difference of two polarizations of polarized light in a magnetic
field. The absorption of photons in vacuum hints that some particles
light enough (milli-eV range) are produced in the apparatus. The
question is what those light particles could be. One well-known
possibility is an axion-like particle ($a$) production in the
magnetic field by the Primakoff interaction $ {\cal
L}_{a\gamma\gamma}=\frac{a}{8M}\,
\epsilon_{\mu\nu\rho\sigma}F^{\mu\nu} {F}^{\rho\sigma}$ where $ 1\
{\rm meV} \lesssim m_a \lesssim 1.5\ {\rm meV} $ and the $a$
coupling to photon in the region $ 2 \times 10^5\ {\rm GeV} \lesssim
M \lesssim 6\times 10^5\ {\rm GeV}.$ The Primakoff process is shown
in Fig. \ref{fig:FeynPVLAS}(a). However, the bound on $M$ is in
direct contradiction with the CAST axion search result, constraining
the axion decay parameter in the range $M>10^{10}$ GeV
\cite{CASTexp}.

 This leads us to consider another light particles $f$
and $\overline{f}$ which can be produced by the photons as shown in
Fig. \ref{fig:FeynPVLAS}(b) and $f$ is called $\lq$millicharged'
fermion~\cite{Gies:2006ca}. For this process to be possible, we must
resolve two problems, how both the $f$ mass and the photon coupling
to $f$ can be so small,
\begin{align}
 &m_f\simeq 0.1\ {\rm eV},\label{eq:massf}\\
& \epsilon_f\equiv Q_f/|e| \simeq 3\times 10^{-6},\label{eq:Qfrac}
\end{align}
where $-e$ is the electron electric charge.

All observed integer electric charges of the known color singlet
particles strongly suggest the quantization of the electric charges
of all elementary particles. The experimental upper limits on the
violation of electric charge quantization was obtained from the
experiments on neutrons \cite{Marinelli:1983nd}, atoms
\cite{Dylla:1973}, and molecules \cite{Baumann:1988ue}: $Q/e <
{\mathcal O}(10^{-21})$. If we introduce the millicharged particle
$f$ with (\ref{eq:Qfrac}) in the standard model (SM), then all
electromagnetic charges are integer multiples of $\epsilon_fe$ if
magnetic monopoles exist \cite{Dirac1931}. Thus a millicharged
particle at the fundamental level introduces a severe problem in the
SM.

But going beyond the SM, there exists a possibility of introducing
such a small unquantized charge, by introducing an additional
hidden-sector U(1) factor \cite{Okun1982}. Holdom noted that the
millicharge is not in conflict with the charge quantization if it
gets an induced electric charge proportional to some small mixing
 between the kinetic terms of photons and extra (or exotic)
photons \cite{Holdom1985}, {\it exphotons}. For the electromagnetic
U(1)$_{\rm em}$ and at least one more hidden sector U(1)$_b'$
factor, the kinetic mixing is usually parametrized as
\begin{equation}
{\mathcal L}_{\rm KE} = -\frac{1}{4} F^{\mu\nu}_{({\rm em})}
F_{({\rm em})\mu\nu} -\frac{1}{4} F^{'\mu\nu}_{(b)} F'_{(b)\mu\nu}
+\frac{\chi}{2} F^{\mu\nu}_{({\rm em})}
F'_{(b)\mu\nu},\label{eq:kinmix}
\end{equation}
where $\chi$ parametrizes the mixing. The  parameter $\chi$ is
directly linked to charge shifts. Eq.(\ref{eq:kinmix}) is written in
the basis where the interaction terms have the canonical form, i.e.
here the electric charge quantization is manifest. For example, with
two fermion species $f_a$ and $f_b$ with charges $(e,0)$ and
$(0,e)$, respectively, under ${\rm U}(1)_{\rm em}\times {\rm
U}(1)_b'$, after diagonalization of ${\mathcal L}_{\rm KE}$, their
charges are known to be shifted by $\epsilon_f \simeq \chi$ in the
leading order of $\chi$ \cite{Holdom1985}.

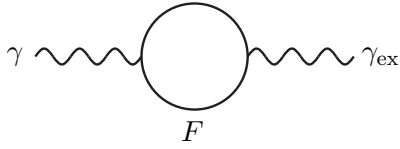
\begin{figure}[t]
\begin{center}
\begin{picture}(400,40)(0,0)
\SetWidth{0.9} \Photon(140,20)(180,20){3}{3} \CArc(200,20)(20,0,360)
\Photon(220,20)(260,20){3}{3} \Text(136,20)[r]{$\gamma$}
 \Text(264,20)[l]{$\gamma_{\rm ex}$}
  \Text(200,-8)[c]{$F$}
\end{picture}
\caption{Mixing of  U(1)$_{\rm em}$ and U(1)$_{\rm ex}$ gauge bosons
through a heavy particle $F$.}\label{fig:kinmixing}
\end{center}
\end{figure}

However, it was argued that introducing an exphoton with millicharge
(\ref{eq:Qfrac}) alone is not free from all astrophysical
constraints. Stellar cores can lose energy much more efficiently by
plasma production of $f$ and $\overline{f}$ compared to the energy
loss via weak interactions. The current astrophysical bound on
$\epsilon_f$ is $\epsilon\lesssim\, 2\times 10^{-14}$ for
$m_f\,\lesssim$\,few keV \cite{HBcore}, which is eight orders of
magnitude below the one needed for the interpretation of the PVLAS
data. This confrontation has led to a need for introducing at least
two extra U(1)$'_b$s in the hidden sector for the case of Fig.
\ref{fig:FeynPVLAS}(b) \cite{Masso2006} and a dimension 6 coupling
without the dimension 5 coupling for the case of
\ref{fig:FeynPVLAS}(a) \cite{Moh06}. In particular, Masso and
Redondo \cite{Masso2006} considered an exactly massless exphoton and
another light exphoton with its mass in the range, keV~$\gg
m_{{\tilde \gamma}}\neq 0$ where $m_{{\tilde \gamma}}$ is the
exphoton plasmon mass in the core of stars. The specific model
considered in \cite{Masso2006} employs a bi-fundamental
representation of {\it two extra} U(1)$'$s, U(1)$'_1$ and U(1)$'_2$,
and obtain the needed condition that $\chi_1=\chi_2$ which is a
fine-tuning condition. In a realistic model from superstring,
however, it is not likely that the conditions as suggested in Ref.
\cite{Masso2006} are satisfied. In Fig. \ref{fig:kinmixing}, a
typical diagram contributing to the kinetic mixing between photon
and U(1)$'_{\rm ex}$ gauge boson is shown.
The threshold correction by heavy particles above and near the GUT
scale for U(1)$'_{\rm em}$ and U(1)$_i$ gauge boson is
 $ \chi_i\simeq -\frac{ee_i}{16\pi^2}\sum_F Q_{\rm em}(F)Q_i(F)
 \ln\frac{{M}_{F}^2}{\mu^2} $ where $e$ is the positron charge,
$e_i$ is the U(1)$_i$ charge, and $\mu$ is the renormalization
scale. Because the heavy particle masses are not identical and
$\sum_F Q_{\rm em}(F)Q_{1}(F)\ne \sum_F Q_{\rm em}(F)Q_{2}(F)$ in
general, the Masso-Redondo condition is difficult to be realized.

In this paper, we introduce just {\it one} U(1)$_{\rm ex}$ gauge
symmetry and a {\it temperature dependent} U(1)$_{\rm ex}$ gauge
symmetry breaking to incorporate  the PVLAS data, the astrophysical
and cosmological bounds. The idea is the following.

We anticipate that the PVLAS experiment has detected the kinetic
mixing effect {\it a la} Fig. \ref{fig:FeynPVLAS}(b). So, suppose
that two U(1) gauge groups survive below the electroweak scale, say
U(1)$_{\rm em}$ and U(1)$_{\rm ex}$. The standard model fermions are
neutral under U(1)$_{\rm ex}$ and the fermions having nonvanishing
U(1)$_{\rm ex}$ charges are called exotics. There exists the mixing
between U(1)$_{\rm em}$ and U(1)$_{\rm ex}$ as shown in Fig.
\ref{fig:kinmixing} through exotic fermions $F$. Suppose that the
mixing parameter $\chi$ turns out to be O($10^{-6}$) to explain the
PVLAS data. U(1)$_{\rm em}$ and U(1)$_{\rm ex}$ being good gauge
symmetries below the electroweak scale, we can break one linear
combination of them and still preserve the remaining combination
$\tilde{\rm U}(1)_{\rm em}$ as the U(1) of QED. There must be a very
light vectorlike exotic fermionic particle pair $f$ and $\bar{f}$
with neutral $\tilde{Q}_{\rm em}$ charge so that their charges after
diagonalization of the kinetic terms is of order $3\times 10^{-6}$.
Without confusion, we will use $f$ to represent the Dirac particle,
$f$ plus $\bar f$. This $f$\ is supposed to be the ones produced at
PVLAS. But its original \Qem\ charge is nonzero. The spontaneous
breaking of U(1)$_{\rm em}$ and U(1)$_{\rm ex}$ is achieved by a
vacuum expectation value (VEV) of a scalar field $\phi$ at a scale
somewhat below keV. The potential of $\phi$ is such that its VEV
vanishes above the critical temperature which is below keV. Thus, at
 stars where $T\sim 2$ keV, the VEV is zero and the original
U(1)$_{\rm em}$ is not broken. Then, in the core of a star $f$\
interacts with photon with the electroweak strength and hence they
are trapped in the star. Trapping of $f$ in the star is like
photons' difficulty in escaping the star.

To formulate the above idea let there be a vectorlike heavy
superfields $F, \bar F$ whose U(1)$_{\rm em}\times$U(1)$_{\rm ex}$
charges are $(\frac13,\frac23)$ and  $(-\frac13,-\frac23)$. The
fractional charges are used to mimick  some singlet exotics of Ref.
\cite{KKK07}. We chose the \Qem\ charge such that it is called
exotics above the VEV scale of $\phi$. The charges of scalar $\phi$
are taken to be $(\frac13,-\frac13)$. Also, the charges of $f$ and
$\bar{f}$ are chosen as $(-\frac13,\frac13)$ and
$(\frac13,-\frac13)$, respectively. Below the critical temperature
$T_{c1}\ll$ keV,  U(1)$_{\rm em}\times$U(1)$_{\rm ex}$ symmetry is
broken down to $\tilde{\rm U}(1)_{\rm em}$. The $\tilde{\rm
U}(1)_{\rm em}$ charges of $f$\ and $\bar{f}$ are zero. Let the
gauge bosons of U(1)$_{\rm em}$ and U(1)$_{\rm ex}$ are $A_\mu$ and
$B_\mu$, respectively, whose coupling to $\phi$ is
 $$
 iD_\mu \phi\to(eA_\mu Q_{\rm em}+e_{\rm ex}B_\mu Q_{\rm ex})\phi=
 \textstyle\frac13(eA_\mu -e_{\rm ex}B_\mu)\phi.
 $$
Thus the photon $\gamma_\mu$ (the gauge boson of QED) and the
sub-keV exphoton $E_\mu$ are
\begin{equation}
\gamma_\mu=\cos\theta A_\mu +\sin\theta B_\mu,\quad E_\mu=
-\sin\theta A_\mu +\cos\theta B_\mu
\end{equation}
where $\tan\theta=e/e_{\rm ex}$. Because the generic value of $e/
e_{\rm ex}$ is not 1 at the electroweak scale, we have generically
$\cos^2\theta\ne\sin^2\theta$. Thus, the kinetic mixing in terms of
the redefined fields is
\begin{equation}
\textstyle -\frac14(1+2\chi\sin\theta\cos\theta)
\gamma_{\mu\nu}\gamma^{\mu\nu}
 -\frac14 (1-2\chi\sin\theta\cos\theta)E_{\mu\nu}E^{\mu\nu}
  -\frac{\chi}{2}(\cos^2\theta-\sin^2\theta) \gamma_{\mu\nu}E^{\mu\nu}.
\end{equation}
Since $\chi$ is small, we can consider a new parameter
$\tilde\chi=\chi(\cos^2\theta-\sin^2\theta)$ for the mixing
parameter of $\gamma_\mu$ and $E_\mu$. It will be again of order
$10^{-6}$. The particles $f$ and $\bar{f}$ have the QED charge 0 but
nonvanishing $E_\mu$ charges $\pm\frac13\sqrt{e^2+e_{\rm ex}^2}$.
The kinetic mixing shifts the QED charge of $f$ from neutral point
by the amount $\epsilon_f=2\tilde\chi/3\sin\theta=
\frac23\chi(\cos\theta\cot\theta -\sin\theta)$. These are the
millicharged particles for the PVLAS experiment. The millicharged
particles of $\tilde{Q}_{\rm em}\sim 3\times 10^{-6}$ are allowed
from all particle physics experiments \cite{Gies:2006ca}.

We introduce a temperature dependent potential such that at the
cores of stars the gauge symmetry is restored \cite{Dolan}. The
temperature dependence regarding Fig. \ref{fig:FeynPVLAS}(a) was
used in Ref. \cite{Moh06}. Thus, at the cores of stars, the original
\Qem s, $-\frac13$ and $\frac13$, of $f$ and $\bar{f}$ are seen
fully and they cannot take out the core energy efficiently. For
them, it is as difficult as photons have the difficulty in escaping
stars. However, they can affect the Big Bang nucleosynthesis (BBN).
The model suggested above has a few new light particles: the gauge
boson $E_\mu$, the complex boson $\phi$, the chiral fermions $f$\
and $\bar{f}$ (or the Dirac particle $f$\ for short). If they are in
equilibrium at BBN, their effective additional cosmologically
equivalent neutrino number ($\delta N_\nu$) is
$\frac87(1+1)+1+1\simeq 4.3$ which is too  large \cite{BBNnew}. So
it is necessary to break the gauge symmetry U(1)$_{\rm ex}$ at the
BBN temperature of order MeV. Then, at the BBN era $f$ and $\bar f$
are milli-charged and they are not in equilibrium with photons and
their number density can be sufficiently lowered assuming that they
are decoupled at a sufficiently early time. The gauge boson $E_\mu$
is heavy at the BBN era and can be neglected. The complex boson
$\phi$ has the vanishing $\tilde Q_{\rm em}$ charge at the
U(1)$_{\rm ex}$ broken phase, and hence it is not in equilibrium
with photon at the BBN era and hence its contribution to $\delta
N_\nu$ can be neglected.

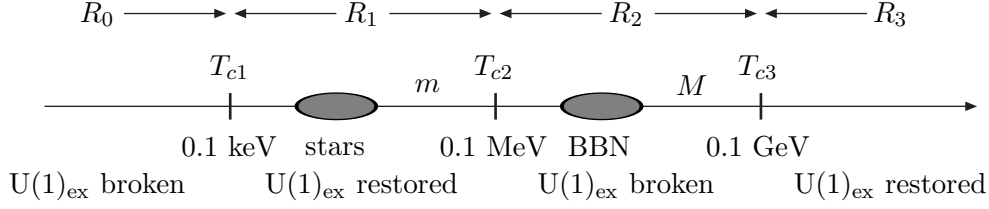
\begin{figure}[h]
\begin{center}
\begin{picture}(400,60)(0,0)

\Text(50,55)[c]{$R_0$}\LongArrow(60,55)(97,55)
\LongArrow(140,55)(103,55)
\Text(150,55)[c]{$R_1$}\LongArrow(160,55)(197,55)
\LongArrow(240,55)(203,55)
\Text(250,55)[c]{$R_2$}\LongArrow(260,55)(297,55)
\Text(350,55)[c]{$R_3$}\LongArrow(340,55)(303,55)

\Text(100,35)[c]{$T_{c1}$} {\SetWidth{1}\Line(100,15)(100,25)}
\Text(200,35)[c]{$T_{c2}$} {\SetWidth{1}\Line(200,15)(200,25)}
\Text(300,35)[c]{$T_{c3}$} {\SetWidth{1}\Line(300,15)(300,25)}
\LongArrow(30,20)(380,20) \Text(100,5)[c]{0.1 keV}
\Text(200,5)[c]{0.1 MeV}  \Text(300,5)[c]{0.1 GeV}

\Text(175,28)[c]{$m$} \GOval(140,20)(5,15)(0){0.5}
\Text(140,5)[c]{stars} \Text(275,28)[c]{$M$}
\GOval(240,20)(5,15)(0){0.5} \Text(240,5)[c]{BBN}

\Text(50,-10)[c]{U(1)$_{\rm ex}$ broken}
\Text(150,-10)[c]{U(1)$_{\rm ex}$ restored}
\Text(250,-10)[c]{U(1)$_{\rm ex}$ broken}
\Text(350,-10)[c]{U(1)$_{\rm ex}$ restored}

\end{picture}
\end{center}
\caption{Typical scales of masses and critical
temperatures.}\label{fig:Tc}
\end{figure}

To have a successful temperature dependent symmetry breaking
pattern, let us introduce the minimal setup with two more real
scalars $\rho$ and $\sigma$. The mass $m$ of $\sigma$  is between
MeV and keV and the mass $M$ of $\rho$ is somewhat above the MeV
scale. The potential of scalar fields $\phi,\rho,$ and $\sigma$ is
\begin{align}
V=\textstyle -\mu^2\phi^*\phi+\frac12 m^2\sigma^2+\frac12 M^2\rho^2
&+\lambda_1(\phi^*\phi)^2-\lambda_2\phi^*\phi\sigma^2
+\lambda_3\phi^*\phi\rho^2+\cdots\label{SSBmodel}
\end{align}
where all parameters are taken to be positive, $\mu^2>0$, $m^2>0$,
$M^2>0$, and $\lambda_i>0\ (i=1,2,3)$, and $\cdots$ represents the
quartic terms of $\rho$ and $\sigma$.  The temperature dependent
potential depending on $\phi$ is estimated as
\cite{Dolan,MS79,comment}
\begin{equation}
V_T=\textstyle\frac12
T^2\left[\lambda_1-\frac12\lambda_2\theta(T-m)+\frac12\lambda_3
\theta(T-M)\right]\phi^*\phi
\end{equation}
where we used the unit Boltzmann constant convention, $k=1$, and
$\theta(x)$ is the step function of $x$.  Then, at zero temperature
the gauge symmetry U(1)$_{\rm ex}$ is broken by the VEV of $\phi$:
$\langle\phi\rangle=v/\sqrt2$, where $v=\sqrt{\mu^2/\lambda_1}$. We
choose a quartic coupling hierarchy as
\begin{equation}
\textstyle\lambda_1\gg \frac12\lambda_2-\lambda_1\gg
\lambda_1-\frac12\lambda_2+\frac12\lambda_3>0.\label{lambdahier}
\end{equation}
Then, there exist three critical temperatures which separate broken
and unbroken phases of U(1)$_{\rm ex}$:
\def\Tca{{$T_{c1}$}}
 \def\Tcb{$T_{c2}$}
 \def\Tcc{$T_{c3}$}
\def\Ro{$R_0$}
\def\Ra{$R_1$}
\def\Rb{$R_2$}
\def\Rc{$R_3$}
\begin{align}
T_{c1}=\textstyle\sqrt{\frac{2\mu^2}{\lambda_1}},&\quad
T_{c2}=\textstyle\sqrt{\frac{4\mu^2}{\lambda_2-2\lambda_1}},\quad
T_{c3}=\textstyle\sqrt{\frac{4\mu^2}{\lambda_3
-\lambda_2+2\lambda_1}},\quad T_{c1}\ll T_{c2}\ll
T_{c3}.\label{crtemps}
\end{align}
There are four phase regions
\begin{equation}
\begin{array}{l}
R_0(T<T_{c1}):\ {\rm U(1)_{ex}\ broken},\quad Q_{\rm em}(f)=\tilde\chi\\
R_1(T_{c1}<T<T_{c2}):\ {\rm U(1)_{ex}\ unbroken},
\quad Q_{\rm em}(f)=\frac13\\
R_2(T_{c2}<T<T_{c3}):\ {\rm U(1)_{ex}\ broken},
\quad Q_{\rm em}(f)=\tilde\chi\\
R_3(T>T_{c3}):\ {\rm U(1)_{ex}\ unbroken},\quad Q_{\rm em}(f)=\frac13\\
\end{array}
\end{equation}
The hierarchy of masses and critical temperatures is shown in Fig.
\ref{fig:Tc}.
The critical temperature \Tca\ is somewhat below 1 keV the typical
temperature in astrophysical environments. The critical temperature
\Tcb\ is somewhat below the MeV scale but above a few keV, and the
critical temperature \Tcc\ is above a few MeV but below the
electroweak scale. These critical temperatures are taken such that
during the \Ro\ period U(1)$_{\rm ex}$ is broken with millicharged
$f$, during the \Ra\ period U(1)$_{\rm ex}$ is restored such that
$f$s are in equilibrium with photons, during the \Rb\ period
U(1)$_{\rm ex}$ is broken again with milli-charged $f$, and during
\Rc\ period U(1)$_{\rm ex}$ is restored such that $f$s are in
equilibrium with photons again. \Ro\ contains the PVLAS region, \Ra\
contains the astrophysical environment and \Rb\ contains the BBN
era. This hierarchy of critical temperatures given in
(\ref{lambdahier}) and (\ref{crtemps}) are achieved by a fine-tuning
of parameters between $\lambda_1,\lambda_2$ and $\lambda_3$.

The spontaneous symmetry breaking of the model in Eq.
(\ref{SSBmodel}) introduces two more real scalars, $\sigma$ and
$\rho$. At the BBN era, $\rho$ is considered to be much heavier than
1 MeV and its contribution to $\delta N_\nu$ can be neglected. [By
the $\lambda_3$ coupling it is in equilibrium with $\phi$ and its
number density is suppressed by the factor $e^{-M/{\rm MeV}}$
compared to that of $\phi$.] So at 1 MeV, we consider the real
scalar $\sigma$, the real part (the Higgs boson of U(1)$_{\rm ex}$)
of the complex scalar $\phi$ and the massive gauge boson $E_\mu$. If
their number densities are the same as the photon number density,
then $\delta N_\nu$ would be $\frac{20}{7}$. But their masses are of
order the temperature scale and we expect that there is a
suppression factor of $e^{-1}$. Thus, we estimate $\delta N_\nu\sim
1$ which is in the allowed region \cite{BBNnew}.

Now let us show briefly how the $\Z_{12-I}$  model of Ref.
\cite{KKK07} contains the needed U(1)$_{\rm ex}$ group and $f$ and
$\phi$. The orbifold compactification leads to the following gauge
group
\begin{equation}
SU(3)_c\times SU(2)\times U(1)^5\times[SO(10) \times U(1)^3]'
\end{equation}
where the electroweak hypercharge $Y$ is $
Y=\textstyle(\frac13~\frac13~\frac13~\frac{-1}{2}~
\frac{-1}{2}~;0~0~0)(0^8)' $ in the E$_8\times$E$_8'$ space and we
obtain $\sin^2\theta_W=\frac38$ \cite{KKK07}. With this hypercharge
assignment, we have exotics: color exotics, doublet exotics, and
singlet exotics. $f,\bar f,$ and $\phi$ belong to singlet exotics
which can be two $\eta$s and one $\overline{\eta}$ exotics of Ref.
\cite{KKK07}.  For example, we can assign $f$ and $\bar f$ as
$\eta_1$ and $\overline{\eta_6}$, respectively. The U(1)$_{\rm ex}$
quantum number is the third entry in the hidden sector
$(\cdot~\cdot~\times;0^5)'$. All the exotics of Ref. \cite{KKK07}
carry nonzero U(1)$_{\rm ex}$ charges, and all non-exotics carry the
zero U(1)$_{\rm ex}$ charge. With the above hypercharge $Y$,
U(1)$_{\rm ex}$ gauge boson {\it exphoton} is massless. From
\cite{KKK07}, the sum of products of the charges, excluding those of
$f, \bar f$ and $\phi$, is found as $
  \sum_{i} Q_{\rm em}(i)Q_{\rm
ex}(i)=\textstyle\frac{35}{9}e
 e_{\rm ex}
$
where $e$ is the positron charge and $e_{\rm ex}$ is the unit charge
of exotics.

To estimate $\chi$, let us consider a toy model with two chiral
superfields of charges $(Q_a,Q_b)$ and $(Q_a,-Q_b)$ and masses $m$
and $m'$ respectively, their joint contribution to $\chi$ has the
form \cite{Dienes1996},
 \begin{equation}
\chi~=~-\frac{g_ag_b}{16\pi^2}\,Q_aQ_b\,
\log\left(\frac{m^2}{m'^2}\right)~. \label{eq:toymodel}
 \end{equation}
From the exotics of Ref. \cite{KKK07}, the contribution to $\chi$ is
not negligible. On general grounds, its size is estimated as
 \cite{Dienes1996}, $10^{-3}< \chi < 10^{-2}, $ where
$\alpha_Y(M_{\rm GUT})\sim \frac{1}{60}-\frac{1}{25}$ is used. It is
possible to reduce $\chi$ to $O( 10^{-6})$ by mass parameters of the
exotics and the RG running. This kind of reduction is necessary even
if we obtained $\sum_{i} Q_{\rm em}(i)Q_{\rm ex}(i)=0$ due to the
appearance of logarithms in Eq. (\ref{eq:toymodel}). Thus, to derive
the millicharged particle scenario, we need two kinds of fine
tunings, one in the quartic coupling constants
$\lambda_1,\lambda_2,\lambda_3$ and the other in the mass parameters
of heavy exotic particles.

In conclusion, the millicharged particles whose properties have
nontrivial temperature dependence can explain the PVLAS data which
was unexpected from conventional models. As shown here and in other
references \cite{Masso2006,Moh06}, it is very unnatural if not
impossible to explain the PVLAS data with field theory models. The
conclusions of these unnatural models, however, lead to far-reaching
consequences on their origins and a crucial clue to new physics
beyond the SM might be obtained by confirming its validity.

\acknowledgments{I thank K. Dienes, I.-W. Kim and B. Kyae for
correspondences and numerous discussions. This work is supported in
part by the KRF ABRL Grant No. R14-2003-012-01001-0. J.E.K. is also
supported in part by the KRF grants, No. R02-2004-000-10149-0 and
No. KRF-2005-084-C00001.
 }



\end{document}